\documentclass[aps,prl,twocolumn,showpacs,groupedaddress]{revtex4}
\usepackage{amsfonts}
\usepackage{amsmath}
\usepackage{amssymb}
\usepackage{graphicx}

\newcommand{\lsmo}{La$_{0.8}$Sr$_{0.2}$MnO$_3$}

\begin{document}
\title{Angular dependence of the Hall effect of \lsmo\ films}
\date{\today }

\author{Netanel Naftalis$^\dagger$}
\email{naftaln@mail.biu.ac.il}
\author{Noam Haham$^\dagger$}
\affiliation{Department of Physics, Nano-magnetism Research Center, Institute of Nanotechnology and Advanced Materials, Bar-Ilan University,
Ramat-Gan 52900, Israel}
\author{Jason Hoffman}
\author{Matthew S. J. Marshall}
\author{C. H. Ahn}
\affiliation{Department of Applied Physics, Yale University, New Haven, CT 06520-8284}
\author{Lior Klein}
\affiliation{Department of Physics, Nano-magnetism Research Center, Institute of Nanotechnology and Advanced Materials, Bar-Ilan University, Ramat-Gan 52900, Israel}

\begin{abstract}
We find that the Hall effect resistivity ($\rho_{xy}$) of thin films of \lsmo\ varies as a function of the angle $\theta$ between the applied magnetic field and the film normal as $\rho_{xy}=a\cos \theta + b\cos 3\theta$, where $|b|$ increases with increasing temperature and decreases with increasing magnetic field. We find that the angular dependence of the longitudinal resistivity and the magnetization cannot fully explain the surprising term $b$, suggesting it is a manifestation of an intrinsic transport property.

\end{abstract}


\maketitle
\setcounter{secnumdepth}{2}
\section{Introduction}

The Hall effect in magnetic conductors has two contributions: a contribution known as the ordinary Hall effect (OHE) associated with Lorentz force and linked to the presence of a magnetic field $\textbf{B}$, and a contribution known as the anomalous Hall effect (AHE) \cite{AHE} linked to magnetization $\textbf{M}$. The AHE has been variously attributed to spin sensitive scattering, which makes its study relevant to spintronics and to topological features of the conduction bands, which have attracted considerable interest in the context of the quantum Hall effect and topological insulators.

The AHE resistivity, $\rho_{xy}^{AHE}$, is phenomenologically described as $\rho_{xy}^{AHE}=R_s\mu_0M_\bot$, where $M_{\bot}$ is the component of the magnetization perpendicular to the film and $R_s$ is called the AHE coefficient. Extrinsic models relate $R_s$ to the longitudinal resistivity, $\rho_{xx}$, and predict $R_{s}=a_{ss}\rho_{xx}+a_{sj}\rho_{xx}^{2}$, where the linear term in $\rho_{xx}$ is attributed to skew-scattering \cite{RsRho1}, and the quadratic term is attributed to the side-jump mechanism \cite{RsRho2}.

Manganites \cite{review_manganite} known for their colossal magnetoresistance \cite{CMR} are particularly intriguing systems for studying AHE due to strong electron correlations and the existence of multiple competing ground states.  Matl et al. \cite{matl} showed a linear relation between the AHE coefficient and the longitudinal resistivity. On the other hand, Fukumura et al. \cite{bad metal scaling experiment} showed a scaling behavior of the AHE  conductivity $\sigma_{xy} \sim \sigma_{xx}^{1.6-1.8}$ in the hopping conductivity regime of several conductors, including manganites.

Several mechanisms have been suggested to explain the AHE in manganites: a mechanism arising from double exchange quantal phases combined with spin-orbit interaction, which predicts scaling with the reduced magnetization \cite{Lyanda-Geller}, and a real space Berry phase mechanism \cite{ye_skyrmion}, which attributes the AHE to the spatial variation of magnetization induced by skyrmions.

The AHE in manganites has previously been studied with magnetization perpendicular to the film plane. Here, we study the dependence of the AHE in thin films of \lsmo\ (LSMO) on the angle $\theta$ between the magnetic field and the normal to the film plane.
If the Hall effect is determined by the perpendicular components of the magnetic field and the magnetization, a trivial $\cos\theta$ dependence is expected. However, we find that the transverse resistivity ($\rho_{xy}$) follows $\rho_{xy}=a\cos\theta+b\cos3\theta$. We show that the $\cos3\theta$ term is not solely due to the angular dependence of the longitudinal resistivity or the magnetization; therefore, the surprising term is likely a manifestation of an intrinsic transport property that has not been identified so far.

\section{Experiment}

The samples in this study are epitaxial thin films of \lsmo\ grown on single crystal SrTiO$_3$(001) substrates using off-axis RF magnetron sputtering.  Growth was carried out at $660^\circ$C in a process gas of 20$\%$ O$_2$ and 80$\%$ Ar at a pressure of 150 mTorr.  After growth, the samples were cooled to room temperature at a rate of $10^\circ$C per minute in 1 bar of O$_2$.  Film thickness was controlled by deposition time, which was calibrated using ex \emph{situ} x-ray diffraction and x-ray reflectivity measurements.
A 40 nm thick calibration film grown with the same conditions was found to be under tensile strain, with a reduced out-of-plane lattice parameter of 0.385 nm and an in-plane lattice parameter of 0.390 nm.  The rocking curve taken around the 002 Bragg reflection had a full width at half maximum of $0.05^\circ$
The films are patterned to allow transverse and longitudinal resistivity measurements, which are performed with a PPMS-9 system (Quantum Design). The magnetic characterization of the films is performed using an MPMS-XL SQUID magnetometer (Quantum Design).

\section{Results and Discussion}

\begin{figure}
\includegraphics[scale=0.45]{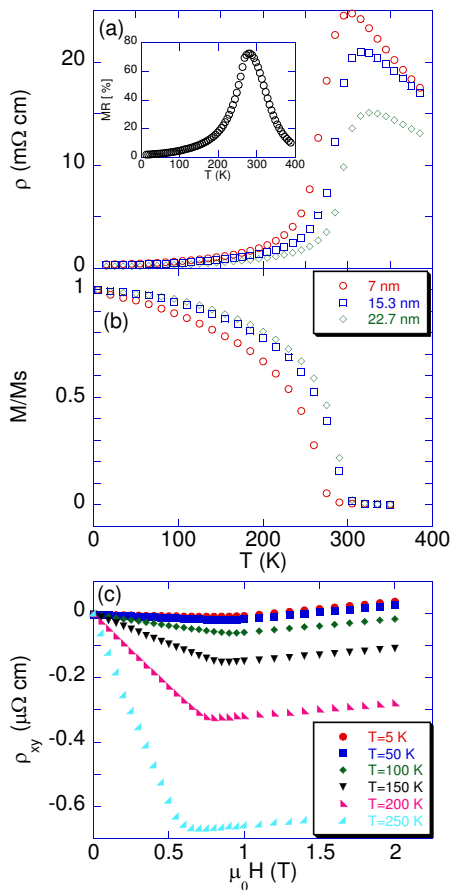}
\caption{(a) Resistivity ($\rho$) as a function of temperature. Inset: Magnetoresistance, $\frac{R_0-R_H}{R_0}*100$, at $\mu_0$H=8 T, as a function of temperature. (b) Reduced magnetization as a function of temperature. (c) $\rho_{xy}$ as a function of magnetic field at different temperatures.}
\label{Fig:MRvsT}
\end{figure}

Figure~\ref{Fig:MRvsT} shows characterization measurements of our films. Figure~\ref{Fig:MRvsT}(a) shows resistivity and magnetoresistance measurements, and Figure~\ref{Fig:MRvsT}(b) shows a field-cooled magnetization measurements with $\mu_0H$=0.05 T. The 22.7 nm thick sample shows bulk-like behavior with a Curie temperature of 300 K and a longitudinal resistivity of 200 $\mu\Omega$-cm at 10 K, pointing to nearly homogeneous electronic and magnetic properties \cite{lsmo_bulk}. Magnetic hysteresis measurements were carried out at 10 K on a 50 nm \lsmo\  film with the field applied perpendicular to the film surface and show the out-of-plane direction to be a hard axis with a saturation magnetization of 3.3 $\mu$B/Mn at fields greater than 2 T.  As shown in Figure~\ref{Fig:MRvsT}(b), the resistivity and Curie temperature depend on film thickness, in agreement with earlier works \cite{thickness}.
Figure~\ref{Fig:MRvsT}(c) shows the magnetic field dependence of the Hall effect resistivity, $\rho_{xy}$, at different temperatures. The two slopes are related to AHE which dominates the change at low fields the OHE which dominates the change at high fields.

\begin{figure}
\includegraphics[scale=0.45]{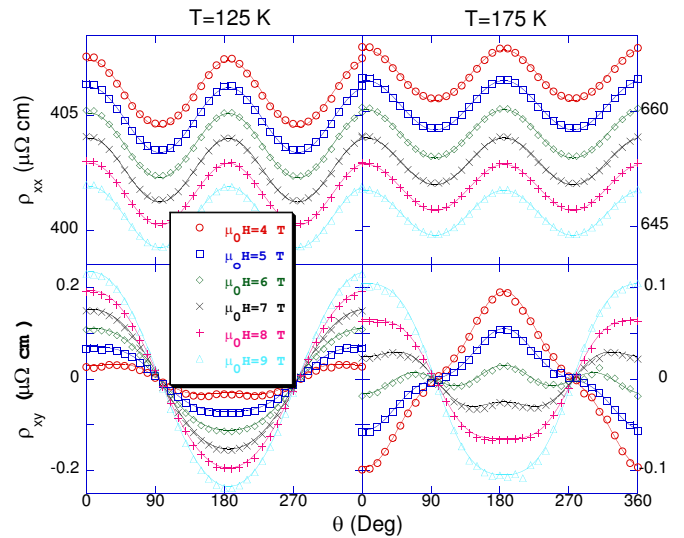}
\caption{$\rho_{xx}$ (top) and $\rho_{xy}$ (bottom) as a function of the external field angle, $\theta$, at T=125 K (left), and T=175 K (right) for different fields. The lines are fits to Eqs. \ref{Eq:AMR} and \ref{Eq:AHE}, respectively.}
\label{Fig:AHE}
\end{figure}

\begin{figure}
\includegraphics[scale=0.5]{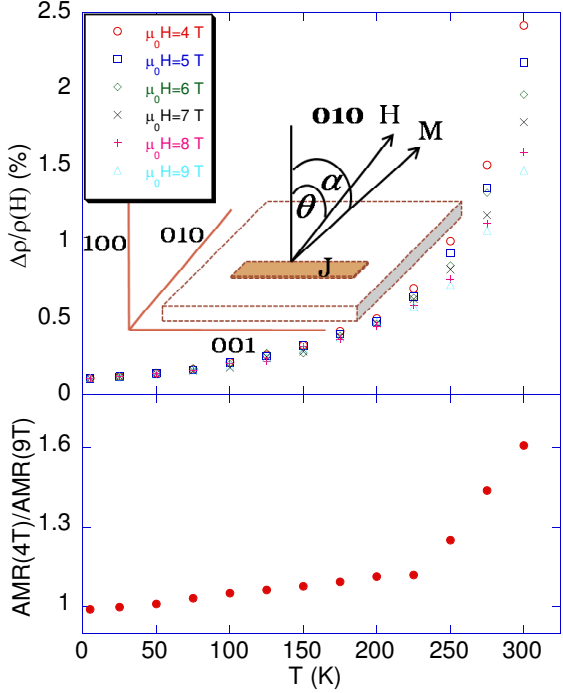}
\caption{(top) $\Delta\rho/\rho$ as function of the temperature for different magnetic fields. Inset: Sketch of the relative orientations of the current density J, magnetic field H, magnetization M, and the crystallographic axes. The angle between H and the film normal, $\theta$, is rotating in the (010) plane. (bottom) The ratio between the AMR amplitude, $\Delta\rho/\rho$, at H=4 and H=9 T, as a function of temperature.}
\label{Fig:AMR}
\end{figure}

Figure~\ref{Fig:AHE} shows $\rho_{xx}$ and $\rho_{xy}$ at ${\rm T=125 \ K}$ and at ${\rm T=175 \ K}$ as a function of the angle $\theta$ between an applied magnetic field ($H$) and the normal to the film (film thickness is 22.7 nm). The current path is along [001], and the field is rotated in the (010) plane (see inset of Figure~\ref{Fig:AMR}). The data are shown for $\mu_0 H$ between ${\rm 4 \ T}$ and  ${\rm 9 \ T}$, for which the magnetization is saturated and parallel to the applied field. We note that the HE does not follow the expected trivial $\cos\theta$ dependence.

The angular dependence of $\rho_{xx}$ is attributed to the anisotropic magnetoresistance (AMR) as follows: $\rho_{xx}=\rho_1+ \rho_2 \cos2\phi$, where $\phi$ is the angle between the current path and the magnetization \cite{AMR}. In the notation we use here:

\begin{equation}
\rho_{xx}=\rho_0+\Delta \rho \cos2\alpha,
\label{Eq:AMR}
\end{equation}
where $\alpha$ is the angle between the magnetization and the film normal.
The good fit in Figure~\ref{Fig:AHE} (top) indicates that the magnetization approximately follows the external magnetic field direction; however, as we will show next, the magnetization has small deviations from the external field direction due to magnetic anisotropy. Figure~\ref{Fig:AMR} (top) shows the AMR amplitude, $\Delta\rho/\rho$, as a function of the temperature for different magnetic fields, and Figure~\ref{Fig:AMR} (bottom) presents the ratio between the AMR measured in two different fields, ${\mu_0 \rm H=4 \ T}$ and ${\mu_0 \rm H=9 \ T}$. We note that below $\sim 200$ K the AMR is practically field independent for $\mu_0 \rm H \geq 4 \ T$, while at higher temperatures the AMR decreases with increasing field in the same range. A decrease in AMR with increase magnetization was observed before and attributed to increased magnetic uniformity \cite{nonmonotonicAMR}.

As noted above, two effects contribute to $\rho_{xy}$: the OHE and the AHE. Commonly, the OHE is proportional to the perpendicular component of the magnetic field, $B_\bot$, and the AHE is proportional to the perpendicular component of the magnetization, $M_\bot$, yielding:

\begin{equation}
\rho_{xy}=R_0B_\bot+R_s\mu_0M_\bot
\label{Eq:HE}
\end{equation}
where $R_0$ and $R_s$ are the OHE and the AHE coefficients, respectively. If $\textbf{M}$ follows the direction of the applied magnetic field and is constant in magnitude, we expect $\rho_{xy}\propto \cos \theta$. The data presented in Fig.~\ref{Fig:AHE}(bottom) clearly deviates from this expectation, whereas a good fit is found with:

\begin{equation}
\rho_{xy}=a\cos \theta + b\cos 3\theta.
\label{Eq:AHE}
\end{equation}

\begin{figure}
\includegraphics[scale=0.5]{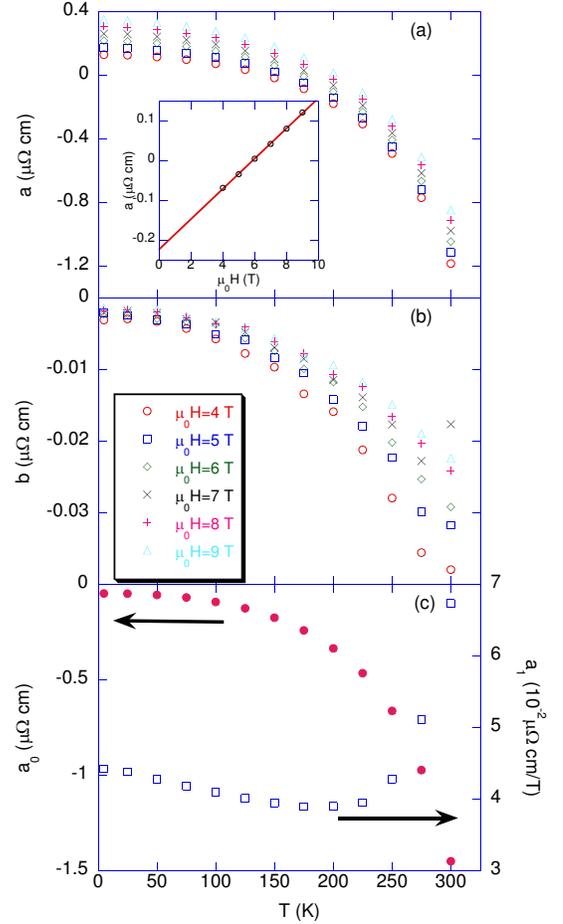}
\caption{The fitting parameters from Eq. \ref{Eq:AHE}, (a) a and (b) b, as function of temperature for different magnetic field. Inset: the field dependence of $a$ at T=175 K. The line is a linear fit. (c) The extrapolated $a_0$ (right) and $a_1$ (right) as function of temperature.}
\label{Fig:parm}
\end{figure}

We fit our data with Eq. \ref{Eq:AHE} in a wide range of temperatures ($5-300$ K) using high magnetic fields ($\mu_{0}H$ between 4 to 9 T). The temperature and field dependence of $a$ and $b$ are shown in Figure~\ref{Fig:parm} for a film thickness of 22.7 nm. A similar behaviour was observed for films with other thicknesses (7 nm and 15.3 nm). We note that at T$\sim$ 150 K the parameter $a$ approaches zero, therefore the contribution of the $\cos 3 \theta$ is more visible.

The parameter $a$ exhibits a linear dependence on the magnetic field, and we denote its slope as $a_1$; the inset of Figure~\ref{Fig:parm}(a) shows this dependence at ${\rm T=175 \ K}$. We extrapolate $a$ to zero and mark it as $a_0$. Figure~\ref{Fig:parm}(c) shows the temperature dependence of the extracted $a_0$ (right) and $a_1$ (left). At low temperatures, where $\textbf{M}$ is close to saturation, we may associate the high-field slope, $a_1$, with the OHE, and $a_0$ with the AHE contribution. This assumption clearly breaks close to $T_c$ where field-induced changes in $\textbf{M}$ may affect the high-field slope of $a$.
The low temperature limit of $a_1$ corresponds to a carrier density of $\sim 1.6\times10^{22}$ carriers per cm$^3$ (i.e., $0.9$ holes per Mn site), larger than the nominal doping level ($\sim 0.2$ holes per Mn site). Similar deviations have been reported before and attributed to the inapplicability of the one-band model \cite{OHE_LSMO,OHE_LCMO}.

\begin{figure}
\includegraphics[scale=0.4]{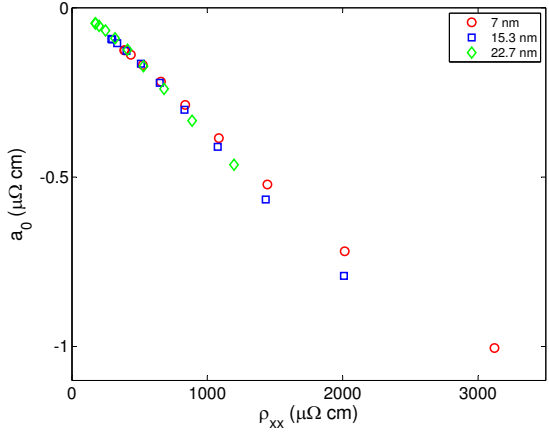}
\caption{The extracted AHE resistivity, $a_0$, as a function of the zero-field longitudinal resistivity.}
\label{Fig:XYvsXX}
\end{figure}

Figure~\ref{Fig:XYvsXX} shows $a_0$ as a function of the zero-field $\rho_{xx}$ for films of different thicknesses. Although the resistivity changes as a function of the thickness, $a_0$ seems to scale with $\rho_{xx}$, consistent with previous reports \cite{matl}. In addition, we note that $a_0\propto\rho_{xx}^{\gamma}$, where $\gamma$ is in the range of 1-1.2.

We turn now to discuss possible sources of the surprising $b$ term.
The OHE would have a contribution with a $\cos 3 \theta$ dependence if in addition to the common term, $R_0 B_\bot$, there would also be a term $R_0^* B_\bot^3$ allowed by symmetry. However, the absolute value of such a term is expected to increase with increasing field, contrary to our observations (see Figure~\ref{Fig:parm}(b)).

Assuming that $b$ is related to the AHE, we consider possible effects of the angular dependence of $\rho_{xx}$ and $M$.
Commonly, the AHE coefficient is described as a function of the longitudinal resistivity; i.e., $R_s=R_s(\rho_{xx})$ \cite{RsRho1,RsRho2,RsRho3,RsRhoNoam}; therefore, one would expect changes in $\rho_{xx}$ to induce changes in the AHE. According to Eq. \ref{Eq:AMR}, $\rho_{xx}$ follows $\cos2\alpha$, which may yield a $\cos3\alpha$ term in the AHE. In other words:
\begin{equation}
\begin{array}{ll}
\rho_{xy}^{AHE}&=R_s(\rho)\mu_0 M_{\bot}\\&=[R_s(\rho_0)+\frac{dR_s}{d\rho}\Delta\rho\cos2\alpha]\mu_0|M|\cos\alpha.
\end{array}
\label{Eq:RsRho}
\end{equation}

\begin{figure}
\includegraphics[scale=0.5]{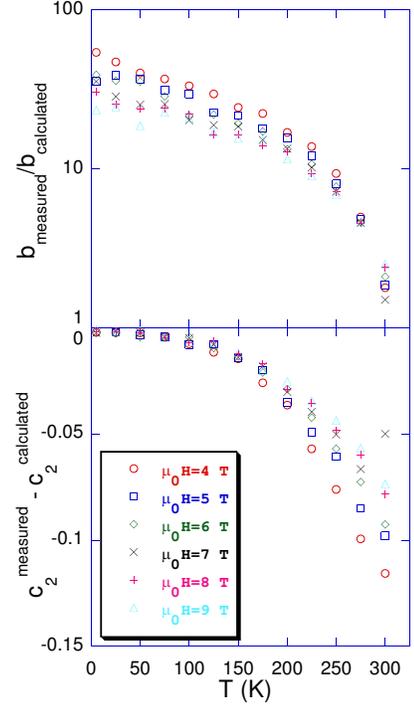}
\caption{(top) The ratio between the measured $b$ to the calculated value of $b$ according to Eq. \ref{Eq:RsRho}, and (bottom) the difference between the measured and calculated $c_2$, according to Eq. \ref{Eq:RsM} as a function of temperature.}
\label{Fig:Ratio}
\end{figure}

However, while the change in $\rho_{xx}$, noted as $\Delta\rho/\rho$, is insensitive to the magnetic field in the low temperature regime (see Figure~\ref{Fig:AMR}), $b$ changes with magnetic field (see Figure~\ref{Fig:parm}(b)). Moreover, Figure~\ref{Fig:Ratio} (top) shows the ratio between the measured $b$ and the calculated $b$ according to Eq. \ref{Eq:RsRho} as a function of temperature. As can be seen, the ratio is larger than $10$ in the low temperature regime.

\begin{figure}
\includegraphics[scale=0.4]{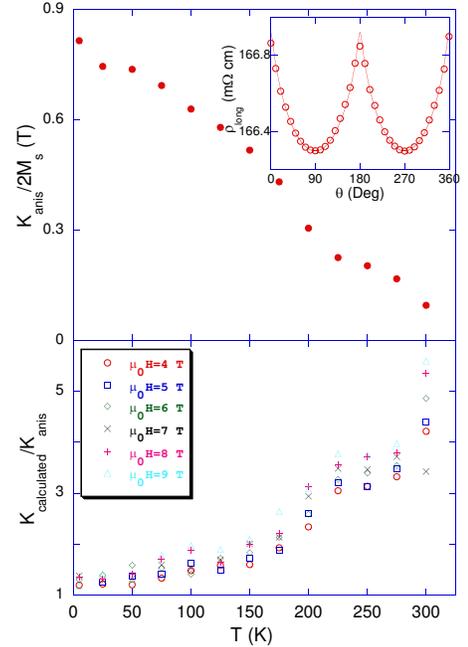}
\caption{(top) $\frac{K_{anis}}{2M_s}$ and (bottom) the ratio between the calculated and measured $K_{anis}$ as a function of temperature. Inset: The angular dependence of the longitudinal resistivity at $\mu_0$H=1 T and T=5 K. The solid line is fit to Eq. \ref{Eq:AMR}, where $\alpha$ is calculated assuming $\frac{K_{anis}}{2M_s}=$0.8 T.}
\label{Fig:Anisotropy}
\end{figure}

Another possible source for $b$ is the angular dependent variation in the magnetization, both in its direction and its magnitude, due to an easy plane anisotropy. Although the magnetization is almost field independent for the magnetic fields that we use, an easy plane anisotropy may cause a small change in the magnetization direction, i.e., the AHE will follow $R_s M \cos(\theta +\Delta\theta) \approx R_s M (\cos\theta -\sin\theta\sin\Delta\theta)$. Considering an easy plane anisotropy and a Zeeman term hamiltonian ($\mathcal{H}=-K_{anis}\sin^2\alpha-M\cdot H\cos(\alpha-\theta)$, where $K_{anis}$ is the anisotropy constant), we obtain that for $H \gg K_{anis}/2M_s$ the deviation from the field direction takes the form $\sin\Delta\theta \approx \frac{K_{anis}}{2M_sH} \cos\theta\sin\theta$. Thus the AHE is given by:
\begin{equation}
\rho_{xy}^{AHE} \approx R_s M \cos\theta (1-\frac{K_{anis}}{2M_sH}\sin^2\theta).
\label{Eq:RsM}
\end{equation}
We extract $\frac{K_{anis}}{2M_s}$ (shown in Figure~\ref{Fig:Anisotropy} (top)) by fitting the longitudinal resistivity to Eq. \ref{Eq:AMR} with $\alpha$ calculated using the easy plane Hamiltonian  with $\frac{K_{anis}}{2M_s}$ as a free parameter (as illustrated in the inset of Figure~\ref{Fig:Anisotropy}).
We fit our measurements with $\rho_{xy}^{AHE}=c_1\cos\theta-c_2\cos\theta\sin^2\theta$ and subtract the expected coefficient $c_2$ based on Eq. \ref{Eq:RsM}, substituting the extracted anisotropy constant.
Figure~\ref{Fig:Ratio} (bottom) shows the difference between the measured and the calculated $c_2$. Figure~\ref{Fig:Anisotropy} (bottom) presents the ratio between the anisotropy constant that which should be assumed in order to explain $c_2$ with this scenario and the measured anisotropy constant as a function of temperature. As can be seen, this scenario yields a good description of $c_2$ below 50 K; nevertheless, this source predicts a field dependent anisotropy constant that is significantly higher than the measured one.

In addition to its effect on the magnetization direction, the magnetic anisotropy may affect the magnitude of the magnetization. According to Eq. \ref{Eq:HE}, the AHE is given by $R_s \mu_0 M_\bot$. Since the magnetization is close to saturation and weakly dependent on $H$, we may approximate its field dependence by $M \approx M(H) + \frac{dM}{dH}\Delta H$. Considering the anisotropy effective field $\frac{K_{anis}}{2M_s}$, we obtain that the total field applied in the direction of the magnetization takes the form: $H^*=H-\frac{K_{anis}}{2M_s}\cos\theta$. Thus, the AHE is given by:
\begin{equation*}
\begin{array}{ll}
R_s \mu_0 M_\bot (\theta)& = R_s \mu_0 [M - \frac{dM}{dH}\frac{K}{2M_s}\cos^2\theta]\cos\theta\\
&=R_s \mu_0 [(M-\frac{dM}{dH}\frac{K}{2M_s})+\frac{dM}{dH}\frac{K}{2M_s}\sin^2\theta]\cos\theta.
\end{array}
\end{equation*}
Therefore this source yields a positive contribution to $c_2$, which cannot explain the measured negative $c_2$.

As we have ruled out trivial sources for the $b\cos 3\theta$ term, it appears that other sources should be considered.
We note that as the temperature increases, $|b|$ increases; and as the magnetic field increases, $|b|$ decreases (see Figure~\ref{Fig:parm}(b)). This behavior may suggest that $|b|$ decreases when $\textbf{M}$ approaches saturation either by increasing the field or by decreasing the temperature. It has been shown that spatial variations in the magnetization may yield a contribution to the AHE in manganites \cite{ye_skyrmion}; however, a $\cos{3\theta}$ dependence is not expected in this model. We point out that structural and magnetic symmetries were previously identified as sources for a more complicated angular dependence of the AMR and PHE in epitaxial films of manganites \cite{AMR_manganites}; however, we do not see a simple way to correlate such effects with our observations.

In conclusion, we find that the AHE in \lsmo\ films cannot be described by the simple relation to the perpendicular component of the magnetization. A more careful treatment that takes into account the lattice and magnetic anisotropies is required.

\section{Acknowledgment}
L.K. acknowledge support by the Israel Science Foundation founded by the Israel Academy of Science and Humanities. Work at Yale supported by NSF MRSEC DMR 1119826 and ONR.

\end{document}